\documentclass[11pt,reqno]{amsart}
\usepackage{amsfonts}
\usepackage{mathrsfs}
\usepackage{amsmath}
\usepackage{amssymb,amsfonts}

\newtheorem{thm}{Theorem}[section]

\newtheorem{defn}{Definition}[section]
\newtheorem{prop}{Proposition}[section]
\numberwithin{equation}{section}
\newtheorem{rmk}{Remark}[section]

\def\pf{{\textit {Proof:} }}

\newcommand{\mysection}[1]{\section{#1}\setcounter{equation}{0}}

\newfont{\bb}{msbm10 at 11pt}

\newcommand{\bal}{\begin{aligned}}      \newcommand{\eal}{\end{aligned}}
\newcommand{\ba}{\begin{array}}      \newcommand{\ea}{\end{array}}
\newcommand{\bc}{\begin{center}}     \newcommand{\ec}{\end{center}}
\newcommand{\be}{\begin{enumerate}}  \newcommand{\ee}{\end{enumerate}}
\newcommand{\beq}{\begin{eqnarray}}  \newcommand{\eeq}{\end{eqnarray}}
\newcommand{\beQ}{\begin{eqnarray*}} \newcommand{\eeQ}{\end{eqnarray*}}
\newcommand{\bi}{\begin{itemize}}    \newcommand{\ei}{\end{itemize}}
\newcommand{\bt}{\begin{tabular}}    \newcommand{\et}{\end{tabular}}
\newcommand{\bdm}{\begin{displaymath}} \newcommand{\edm}{\end{displaymath}}

\def\qed{\hfill{Q.E.D.}\smallskip}

\newcommand{\ls}{\setlength{\baselineskip}{12pt}
	\setlength{\parskip}{3mm}}

\begin{document}
	
\allowdisplaybreaks

 \title[Stochastic Radial Motion]{Stochastic motion of massless particles near event horizons in Schwarzschild and Reissner-Nordstr\"om spacetimes}

 \author[Y Huang]{Yinhan Huang$^{1,2}$}
 \author[L Miao]{Lezhen Miao$^{3,5}$}
 \author[X Zhang]{Xiao Zhang$^{1,2,4}$}
 
 \address[]{$^1$Academy of Mathematics and Systems Science, Chinese Academy of Sciences, Beijing 100190, PR China}
 \address[]{$^2$School of Mathematical Sciences, University of Chinese Academy of Sciences, Beijing 100049, PR China}
 \address[]{$^3$School of Mathematics, Guangxi University, Nanning, Guangxi 530004, PR China}
 \address[]{$^4$Guangxi Center for Mathematical Research, Guangxi University, Nanning, Guangxi 530004, PR China}
 \address[]{$^5$Ningde No.5 Middle School of Fujian, Ningde, Fujian 352100, PR China}
 
 \email{(Huang) huangyinhan@amss.ac.cn}
 \email{(Miao) 1257951432@qq.com}
 \email{(Zhang) xzhang@amss.ac.cn}

 \date{}

\begin{abstract}
In classical general relativity, it is well known that radially moving massless particles in the exterior region take infinite coordinate time to reach the event horizon in black hole spacetimes. In this short paper, we use two stochastic versions of ingoing radial null geodesic equation to study the stochastic radial motion of massless particles in Schwarzschild and Reissner-Nordstr\"om spacetimes. We show that the probability that a radially moving massless particle enters the black hole in finite coordinate time remains zero. This indicates the stochastic perturbations of quantum effects preserve certain properties of classical black holes.
\end{abstract}

\maketitle \pagenumbering{arabic}

\mysection{Introduction}\ls

Black holes are among the most important solutions of Einstein field equations. In Schwarzschild spacetime with mass $M>0$
\begin{equation*}
    ds^2=-\Big(1-\frac{2M}{r}\Big)dt^2+\Big(1-\frac{2M}{r}\Big)^{-1} dr ^2+r ^2 \Big( d\theta ^2 +\sin^2 \theta d\phi ^2 \Big),
\end{equation*}
the ingoing radial null geodesics satisfies
\begin{equation}
    dt = -\Big(1+\frac{2M}{r-2M}\Big)dr.\label{Schwarzschild}
\end{equation}
Starting at some point ($t_0$, $r_0$) outside the event horizon ($r_0 > 2M$), the radial trajectory of a massless particle is
\begin{equation*}
	t-t_0=r_0-r-2M\ln(r-2M)+2M\ln(r_0-2M).
\end{equation*}
The above equation directly implies the classical result
\begin{equation*}
	\lim_{r \to 2M} t = +\infty.
\end{equation*}
It takes infinite coordinate time for massless particles in the exterior region to reach the event horizon $r = 2M$.

The same phenomenon also occur in Reissner-Nordstr\"om spacetime with mass $M>0$
\begin{equation*}
    ds^2 = -\Big(1-\frac{2M}{r} + \frac{e^2}{r^2}\Big)dt^2+\Big(1-\frac{2M}{r} +\frac{e^2}{r^2}\Big)^{-1} dr^2+r ^2 \Big( d\theta ^2 +\sin^2 \theta d\phi ^2 \Big).
\end{equation*}
For $M < |e|$, there is no event horizon in Reissner-Nordstr\"om spacetime. 
For $M > |e|$, the event horizon and Cauchy horizon lie at 
\begin{align*}
r_+ = M + \sqrt{M-e^2}, \quad r_- = M - \sqrt{M-e^2} 
\end{align*}
respectively, which is referred to as the non-extreme Reissner-Nordstr\"om black hole. We study massless particles starting from the exterior region and reaching the event horizon. The ingoing radial null geodesics satisfies
\begin{equation}
    dt = -\Big(1+\frac{2Mr-e^2}{r^2-2Mr+e^2}\Big)dr = -\Big( 1+ \frac{2Mr-e^2}{(r-r_+)(r-r_-)} \Big)dr. 
    \label{nonextremeRN}
\end{equation}
Starting at some point ($t_0$, $r_0$) outside the event horizon ($r_0 > r_+$), the radial trajectory of a massless particle is
\begin{equation*}
	t-t_0 = r_0-r -A\ln(r-r_+) +A\ln(r_0-r_+) -B\ln(r-r_-)+ B\ln(r_0-r_-),
\end{equation*}
where coefficients
\begin{equation*}
    A= M + \frac{1}{2}\frac{2M^2-e^2}{\sqrt{M^2-e^2}}, \quad 
   B = M -\frac{1}{2}\frac{2M^2-e^2}{\sqrt{M^2 - e^2}}. 
\end{equation*}
The above equation also implies the classical result
\begin{equation*}
	\lim_{r \to r_+} t = +\infty.
\end{equation*}

For $M = |e|$, the Cauchy and event horizons of Reissner-Nordstr\"om black hole coincide at $r_+ = r_- = M$, which is referred to as the extreme Reissner-Nordstr\"om black hole.
Similar to the above derivation, the ingoing radial null geodesics satisfies
\begin{equation}
    dt = -\Big(1+ \frac{2M}{r-M} + \frac{M^2}{(r-M)^2}\Big)dr. \label{extremeRN}
\end{equation}
The radial trajectory of a massless particle is
\begin{equation*}
	t-t_0 = r_0-r -2M \ln(r-M) + 2M\ln(r_0-M) + \frac{M^2}{r-M} - \frac{M^2}{r_0-M}.
\end{equation*}
Thus
\begin{equation*}
	\lim_{r \to M} t = +\infty.
\end{equation*}

However, some new phenomena may arise if quantum effects of black holes are considered. For instance, within the framework of semiclassical gravity, Hawking\cite{H} showed that black holes emit Hawking radiation and can eventually evaporate. On the other hand, in the deformation quantization approach, the quantum Schwarzschild black hole is constructed, and it does not evaporate, and the area of event horizon is decreased as
\begin{equation*}
	16\pi M^2 \Big(1-\frac{\bar{h}^2}{6} +O(\bar{h} ^4)\Big),
\end{equation*}
where $\bar{h}$ is the Planck constant\cite{CTZZ, WZZ}. 

In stochastic gravity (see \cite{HV} for a review), the quantum fluctuations are described by noise kernels, and the semiclassical Einstein equations are replaced by the Einstein-Langevin equations. It is interesting to investigate how black holes are affected by this kind of quantum fluctuations. In this paper, we show that 
the probability remains zero for massless particles in the exterior region to reach the event horizon in any finite coordinate time.

The paper is organized as follows. In Section 2, we review some background on stochastic process and derive stochastic differential equations for the radial motion of massless particles. In Section 3, we outline the derivation of the Fokker-Planck equation. In Sections 4 and 5, we analyze the characteristics of the probability density function of the radially moving massless particle in Schwarzschild and Reissner-Nordstr\"om spacetimes respectively, and prove the main theorems.

\mysection{Stochastic differential equations of motion}\ls

In this section, we introduce some basic concepts of stochastic process, including Markov chain, Brownian motion and fractional Brownian motion based on \cite{G1,G2,G3}.
\begin{defn}
A stochastic process is a parameterized collection of random variables
$$\{X_t\}_{t\in T}$$
defined on a probability space $(\Omega,\mathscr{F},P)$ and taking values in $\mathbb{R}^n$. 

For each $t \in T$,
$$\omega\to\{X_t\}_{\omega}; \qquad \omega\in\Omega $$
is a random variable. For each $\omega\in\Omega$,
$$ t\to\{X_t\}_{\omega}; \qquad t\in T$$
is called a path of $X_t$.
\end{defn}

It is useful to think of $t$ as time and each $\omega$ as an individual particle or experiment. Within this picture, ${X_t}(\omega)$ would represent the position of the particle $\omega$ at time $t$.

\begin{defn}
Markov process is the stochastic process $\{X_t\}_{t\in T}$ whose conditional probability satisfies
\beQ
\begin{aligned}
	 P\{X(t) &\le x  \mid X(t_n) =x_n,\dots ,X(t_1)=x_1 \}\\
                  &=P\{X(t)\le x\mid X(t_n)=x_n\}
\end{aligned}
\eeQ
for all $t_i\in T$, $t_1<t_2<\dots<t_n<t$.
\end{defn}
\begin{defn}

Brownian motion is the stochastic process $\{{B_t}(\omega), t\ge0\}$ satisfying
\begin{itemize}
	\item[(i)] $B_0=0$,
	\item[(ii)] $B_t$ has independent increments, i.e.
$$B_{t_1}, B_{t_2}-B_{t_1}, \cdots, B_{t_k}-B_{t_{k-1}}$$
are independent for all $0 \leq t_1 < t_2 \cdots <t_k$,
	\item[(iii)] $B_t$ is a Gaussian process, i.e. for all $0 \leq t_1 < t_2 \cdots <t_k$, the random variable $Z = (B_{t_1},\cdots ,B_{t_k}) \in  \mathbb{R}^{nk} $ has a (multi)normal distribution.
\end{itemize}
\end{defn}

From the above definition, Brownian motion is a Markov process. Hence, under the assumption of no long memory, stochastic differential equations driven by Brownian motion can model many stochastic phenomena, including the Black–Scholes option pricing model. However, many natural phenomena exhibit long memory, motivating the search for other stochastic processes. Accordingly, we introduce the following definition of fractional Brownian motion.

\begin{defn}
	A Gaussian process $B^H = \{B^H(t),t\ge 0\} $ is called fractional Brownian motion of Hurst parameter $H \in (0, 1) $ if it has mean zero and the covariance function
    \begin{equation*}
     E[B^H(t)B^H(s)]=\frac{1}{2}\big(t^{2H}+s^{2H}-{|t-s|}^{2H} \big).
    \end{equation*}
\end{defn}

The main difference is that the increment in Brownian motion is independent, while the increment in fractional Brownian motion is not independent (except for $H=\frac{1}{2}$), and it is not a Markov process.

The standard stochastic differential equation driven by Brownian motion can be written as (interpreted in the It\^o sense) 
\begin{equation}
    dX_t = b(t, X_t) dt + \sigma(t, X_t)dB_t.
    \label{SDE}
\end{equation} 

Analogously, a stochastic differential equation driven by fractional Brownian motion is often written in the Wick--It\^o--Skorohod framework (abbreviated as fWIS below)
\begin{equation}
    dX_t = b(t, X_t) dt + \sigma(t, X_t)dB_t^H.
    \label{fSDE}
\end{equation}

\mysection{Fokker-Planck equation}\ls

In this section, we present a brief derivation of the Fokker-Planck equation and fractional Fokker-Planck equation based on \cite{G1, G2, G3}. The derivation presented here is intended mainly for the reader's convenience and to keep the paper self-contained. We refer to \cite{G1, G2, G3} for a more thorough and systematic treatment.
 
We denote $f(x, t)$ as the probability density for the stochastic process $X_t$.
\begin{defn}
    The conditional characteristic function $\phi(u,t \mid x_0,t_0)$ is the conditional expectation of $e^{iux}$ given $x_0,t_0$, 
    \begin{equation*}
        \phi(u,t \mid x_0,t_0) = \mathbb{E}[e^{iuX_t} \mid X_{t_0} = x_0].
    \end{equation*}
    Equivalently, it can be computed as
    \begin{equation*}
        \phi(u,t \mid x_0,t_0) = \int_{-\infty}^{+\infty} e^{iux} f(x,t \mid x_0,t_0) \, dx.
    \end{equation*}
\end{defn}
    Note that the characteristic function of a stochastic process plays the same role as Fourier transform, with the opposite sign convention in the exponent.
    
    The inverse transform is
 \begin{equation}
      \begin{aligned}
 f(x,t+\Delta t|x^\prime,t) =& \mathcal{F}^{-1}[\phi(u,t+\Delta t|x^\prime,t)] \\
 	  	                    =& \frac{1}{2\pi} \int_{-\infty}^{\infty} e^{-iu\Delta x}\phi(u,t+\Delta t|x^\prime,t)\, du, 	  	
 \end{aligned}
 \label{inverse}
 \end{equation}
where $\Delta x=x-x^\prime$.

The Taylor expansion of $\phi$ at $ u=0 $ is
\begin{equation}
    \phi(u,t+\Delta t|x^\prime,t)=\sum_{n=0}^\infty \frac{(iu)^n}{n!}\alpha_n(x^\prime,t)
\label{Taylor}
\end{equation}
where $	\alpha_n(x^\prime,t) $ is called the incremental moment of $n$ order defined by
\begin{equation*}
	\alpha_n(x^\prime,t)=\mathbb{E}\{[X(t+\Delta t)-X(t)]^n |x^\prime,t\}.
\end{equation*}
Combining Eq\eqref{inverse} and Eq\eqref{Taylor}, we have
\begin{equation*}
    \begin{aligned}
	f(x,t+\Delta t|x^\prime,t) =& \sum_{n=0}^\infty \frac{\alpha_n(x^\prime,t)}{2\pi n!}\int_{-\infty}^{\infty} (iu)^n e^{-iu\Delta x}\, du \\
	                           =& \sum_{n=0}^\infty \frac{(-1)^n}{n!} \alpha_n(x^\prime,t) \frac{\partial^n}{\partial x^n}[\delta(\Delta x)].
\end{aligned}
\end{equation*}
The above expansion and term-wise integration are formal. However, for the cases we considered in Sections 4 and 5, the convergence is automatically satisfied.
Note that
\begin{equation*}
    f(x,t+\Delta t)=\int_{-\infty}^{\infty}	f(x,t+\Delta t|x^\prime,t)f(x^\prime,t)\,dx^\prime
\end{equation*}
and
\begin{equation*}
    \int_{-\infty}^{\infty}\delta(x)\,dx=1
\end{equation*}
Thus we have
\begin{equation*}
    \begin{aligned}
	f(x,t+\Delta t) =& \sum_{n=0}^\infty \frac{(-1)^n}{n!}\frac{\partial^n}{\partial x^n}[ \alpha_n(x,t)f(x,t)]  \\
                	=& f(x,t)+\sum_{n=1}^\infty \frac{(-1)^n}{n!}\frac{\partial^n}{\partial x^n}
                	[\alpha_n(x,t)f(x,t)].
\end{aligned}
\end{equation*}

Dividing both sides by $\Delta t$ and taking the limit, we obtain the Kramers-Moyal expansion
\begin{equation}
    \frac{\partial f(x,t)}{\partial t}-\sum_{n=1}^\infty \frac{(-1)^n}{n!}\frac{\partial^n}{\partial x^n}[a_n(x,t)f(x,t)] = 0
\label{Kramers-Moyal}
\end{equation}
where $	a_n(x,t) $ is called the derivative moment in incremental moment of $n$ order defined by
\begin{equation*}
   a_n(x,t)=\lim_{\Delta t \to 0} \frac{\alpha_n(x,t)}{\Delta t}
\end{equation*}

The Markov property itself does not imply that the higher order terms in Kramers-Moyal expansion vanish. However, in the case of It\^o diffusion Eq\eqref{SDE}, Kramers-Moyal expansion truncates at second order. We have
\begin{equation*}
    \frac{\partial f}{\partial t}=-\frac{\partial}{\partial x}[a_1(x,t)f]+\frac{1}{2}\frac{\partial^2}{\partial x^2}[a_2(x,t)f]
\end{equation*}
where
\begin{equation}
    a_i(x,t)=\lim_{\Delta t \to 0} \frac{\mathbb{E}\{[\Delta X]^i |X(t)=x]\}}{\Delta t} \label{a-i}
\end{equation}
for $i=1,2$, and $\Delta X$ can be obtained from Eq\eqref{SDE}.
Thus
\begin{equation*}
    \begin{aligned}
a_1(x,t) =b(x,t), \qquad a_2(x,t) =2D\sigma^2(x,t),
\end{aligned}
\end{equation*}
where $D$ is referred to as the diffusion coefficient.

Finally, we obtain the Fokker-Planck equation associated with Eq\eqref{SDE} is
\begin{equation}
    \frac{\partial f}{\partial t}=-\frac{\partial}{\partial x}[b(x,t)f]+D\frac{\partial^2}{\partial x^2}[\sigma^2(x,t)f].  \label{FK}
\end{equation}

Now we introduce the fractional It\^o formula of the fractional Brownian motion in the following proposition, where $ 0<H<1 $.
\begin{prop}
	Let $h \in L^H(R),H \in (0,1)$, defining
	$$\eta(t)=\int_{0}^{t} h(s)\,dB^H(s).$$
	Let $f(t,x):R_+ \times R\to R \in C^{2,1}(R_+ \times R;R)$, we have
 \begin{equation*}
 	\begin{aligned}
 f(t,\eta(t)) =& f(0,0)+\int_{0}^{t} \frac{\partial f}{\partial s}(s,\eta (s))\,ds+\int_{0}^{t} \frac{\partial f}{\partial x}(s,\eta (s)h(s))\,dB^H(s) \\
               & +\int_{0}^{t} \frac{\partial^2 f}{\partial x^2}(s,\eta (s))\,dh(s) \int_{0}^{s} \phi (s,v)h(v)\,dvds,
\end{aligned}
 \end{equation*}
or alternatively
  \begin{equation}
      \begin{aligned}
 	df(t,\eta(t)) =& \frac{\partial f}{\partial t}(t,\eta (t))\,dt+ \frac{\partial f}{\partial x}(t,\eta (t)h(t))\,dB^H(t) \\
 	               & +\frac{\partial^2 f}{\partial x^2}((t,\eta (t))h(t) \int_{0}^{t} \phi (t,v)h(v)\,dvdt.
 \end{aligned}
 \label{fIto}
  \end{equation}
\end{prop}
We now consider Eq\eqref{fSDE}. By using the fractional It\^o formula Eq\eqref{fIto}, we see that the stochastic process $h(x)$ satisfies:
\begin{equation*}
    \begin{aligned}
	dh(x) =&\Big(\frac{dh(x)}{dx}b(t,x(t))+ \frac{d^2h(x)}{d x^2}\sigma(t,x(t)\Big)\int_{0}^{t} \sigma(s,x(s))\phi (t,s)\,ds \big) dt\\
	       & +\frac{dh(x)}{dx}\sigma(t,x(t))dB^H(t) .
\end{aligned}
\end{equation*}
Taking averages on both sides, we obtain
\begin{equation}
    \mathbb{E}\Big[\frac{dh(x)}{dt}\Big] = \mathbb{E}\Big[\frac{dh(x)}{dx}a(t,x(t))+ \frac{d^2h(x)}{d x^2}b(t,x(t))\int_{0}^{t} b(s,x(s))\phi (t,s)\,ds\Big].
    \label{expectation}
\end{equation}
We also have
\begin{equation*}
    \mathbb{E}\Big[h(x)\Big] =\int_{-\infty}^{\infty} h(x)f(x,t)\,dx,
\end{equation*}
so that the derivative of the mean can be written as
\begin{equation*}
    \mathbb{E}\Big[\frac{dh(x)}{dt}\Big] =\int_{-\infty}^{\infty} h(x)\frac{\partial f(x,t)}{\partial t}\,dx.
\end{equation*}
Integrating by parts and using
\begin{equation*}
     \lim_{x \to \pm\infty } f(x,t) =0,
\end{equation*}
we obtain
\begin{equation*}
\begin{aligned}
	\int_{-\infty}^{\infty}\frac{dh(x)}{dx} a(t,x(t))f(x,t)\,dx
    = & \ a(t,x(t))f(x,t)h(x)\Big | _{-\infty}^{\infty}\\
      & -\int_{-\infty}^{\infty}h(x)\frac{\partial \big(a(t,x(t))f(x,t)\big)}{\partial x}\,dx \\
	= & -\int_{-\infty}^{\infty}h(x)\frac{\partial \big(a(t,x(t))f(x,t)\big)}{\partial x}\,dx.
\end{aligned}
\end{equation*}
Similarly, we have
\begin{equation*}
    \begin{aligned}
	& \int_{-\infty} ^{\infty} \frac{d^2h(x)}{dx^2}b(t,x(t)) \Big(\int_{0}^{t} b(s,x(s))\phi (t,s)\,ds \Big) f(x,t)\,dx  \\
	& = \int_{-\infty}^{\infty} h(x)\frac{\partial^2 }{\partial x^2} \Big[b(t,x(t))\Big(  \int_{0}^{t} b(s,x(s))\phi (t,s)\,ds \Big)  f(x,t)\Big]\,dx.
\end{aligned}
\end{equation*}
Since $h(x)$ above is arbitrary, plugging the derivatives of $h(x)$ into Eq\eqref{expectation}, we get the fractional Fokker-Planck equation:
\begin{equation}
    \begin{aligned}
	& \frac{\partial f(x,t)}{\partial t}  +  \frac{\partial }{\partial x}\Big(b(t,x(t))f(x,t)\Big)\\
	& - D\frac{\partial^2}{\partial x^2} \Big[\sigma(t,x(t))\Big(\int_{0}^{t} \sigma(s,x(s))\phi (t,s)\,ds \Big) f(x,t)\Big]=0.
\end{aligned}
\label{fFK}
\end{equation}
where
\begin{equation*}
    \phi(t,s)=H(2H-1) |s-t|^{2H-2}, \quad  s,t \in \mathbb{R}.
\end{equation*}

\mysection{Stochastic motion of massless particles in Schwarzschild spacetime}\ls

In this section, based on the (fractional) Fokker-Planck equation, we calculate the probability for massless particles in the exterior region to reach the event horizon in finite coordinate time. We show that the probability is still zero.

Eq\eqref{Schwarzschild} is a deterministic model in the classical situation. We treat the stochastic gravitational effects of other objects as noise, which we model by a standard Brownian motion. The corresponding stochastic differential equation (SDE) can be written as
\begin{equation}
    \begin{aligned}
dt=-\Big(1+\frac{2M}{r-2M}\Big)dr+dB(r), \qquad  2M \leq r \leq 3M.         \label{Schwarzschild-B}
\end{aligned}
\end{equation}
This equation can be used to describe the stochastic radial motion of massless particles in Schwarzschild spacetime.
In order to keep the positive range of variable, we change the coordinates as follows
\begin{equation*}
	r=3M-r^*.
\end{equation*}
Substituting it into Eq\eqref{Schwarzschild-B}, we obtain
\begin{equation}
    dt=\Big(1+\frac{2M}{M-r^*}\Big)dr^*+dB(r^*), \qquad  0 \leq r^* \leq M.
    \label{Schwarzschild-Brownian}
\end{equation}
The corresponding fWIS stochastic equation is
\begin{equation}
    dt=\Big(1+\frac{2M}{M-r^*}\Big)dr^*+dB^H(r^*), \qquad  0 \leq r^* \leq M.
    \label{Schwarzschild-fBrownian}
\end{equation}

    Now we derive the probability density functions of the above two motion processes.
\begin{prop}\label{p-Schwarzshild-Brownian} 
 	The probability density function of the stochastic radial motion driven by Brownian motion of massless particles in Schwarzschild spacetime is
 	\begin{equation*}
 	    f(t,r^*|t_0,r^*_0)  = \frac{1}{\sqrt{2\pi \sigma^2_{t(r^*)}}}
 		\exp\Big[-\frac{\big({t-\mu_{t(r^*)} }\big ) ^2}{2\sigma^2_{t(r^*)}}\Big],
 	\end{equation*}
 	where
    \begin{equation*}
 	\mu_{t(r^*)} = t_0-\Big(r^*-r^*_0-2M\ln \frac{r^*-M}{r^*_0-M}\Big), \qquad \sigma^2_{t(r^*)} =  2D (r^*-r^*_0).
    \end{equation*}
\end{prop}

\pf
    Combining Equations \eqref{FK} and \eqref{Schwarzschild-Brownian}, we obtain
   \begin{equation*}
       \frac{\partial f(t,r^*|t_0,r^*_0)}{\partial r^*}=-\Big(1+\frac{2M}{M-r^*}\Big)\frac{\partial f}{\partial t}+D\frac{\partial^2 f}{\partial t^2}.
   \end{equation*}
    The initial condition is
    \begin{equation*}
    f(t,r^*|t_0,r^*_0)=\delta(t-t_0).
    \end{equation*}
    When $t = + \infty$, the particle has entered the black hole region. So the boundary condition is
    \begin{equation*}
        f(\pm \infty,r^*|t_0,r^*_0)=0.
    \end{equation*}
    Consider the characteristic functions (equivalently, take Fourier transform) of both sides and recall
    \begin{equation*}
        \phi(u,r^*|t_0,r^*_0)=\int_{-\infty}^{\infty} e^{iut}f(t,r^*|t_0,r^*_0)\, dt,
    \end{equation*}
we obtain
\begin{equation*}
    \begin{aligned}
     \mathcal{F} \big(\frac{\partial f}{\partial r^*}\big) & = \frac{\partial}{\partial r^*}\int_{-\infty}^{\infty} e^{iut}f\, dt=\frac{\partial \phi}{\partial r^*}, \\
     \mathcal{F} \big(\frac{\partial f}{\partial t}\big)   & = iu \phi, \\
     \mathcal{F} \big(\frac{\partial^2 f}{\partial t^2}\big) & = -u^2 \phi.
    \end{aligned}
\end{equation*}
    Then we get
\begin{equation*}
   \frac{\partial \phi}{\partial r^*}=-\Big(1+\frac{2M}{M-r^*}\Big) iu \phi-Du^2 \phi. 
\end{equation*}  
    Integrate the above equation, we get
      \begin{equation*}
          \phi = C\ \exp\! \Big\{iu  \Big[ 2M\ln \frac{r^*-M}{r^*_0-M}-(r^*-r^*_0)\Big]  -Du^2(r^*-r^*_0)\Big\}.
      \end{equation*}
    The initial condition is
       \begin{equation*}
           \phi(u,r^*_0|t_0,r^*_0) =\int_{-\infty}^{\infty} e^{iut}\delta(t-t_0)\, dt =e^{iut_0}.
       \end{equation*}
      Then we have
      \begin{equation*}
          \phi = \exp\! \Big\{iu  \Big[t_0-\big(r^*-r^*_0-2M\ln \frac{r^*-M}{r^*_0-M}\big)\Big]  -Du^2(r^*-r^*_0)\Big\}.
      \end{equation*}
    Taking the inverse Fourier transform on both sides and we obtain
        \begin{align*}
          f(t,r^*|t_0,r^*_0)
           =& \mathcal{F}^{-1}\Big[ \exp\big( iu  \mu _{t(r^*)} \big)\cdot \exp\big(-Du^2(r^*-r^*_0) \big)  \Big]  \\
                   =& \mathcal{F}^{-1}\Big[  \exp\big( iu  \mu _{t(r^*)} \big) \Big]* \mathcal{F}^{-1}\Big[ \exp\big( -Du^2(r^*-r^*_0) \big)  \Big] \\
                   =&  \delta \big(t-\mu _{t(r^*)}\big) * \frac{1}{2\sqrt{\pi D(r^*-r^*_0)}} \exp \Big[-\frac{t^2}{4D(r^*-r^*_0)} \Big] \\
                   =&  \frac{1}{2\sqrt{\pi D(r^*-r^*_0)}}\exp \Big[-\frac{\big(t-\mu _{t(r^*)}\big)^2}{2\big(\sqrt{2 D(r^*-r^*_0)}\big)^2}\Big].          
         \end{align*}  	  	
        \qed

       Theorem \ref{p-Schwarzshild-Brownian} shows that the stochastic process $t(r^*)$ governed by the linear stochastic differential equation \eqref{Schwarzschild-Brownian} is Gaussian, with expectation $\mu_{t(r^*)}$ and standard deviation $\sigma_{t(r^*)}$. Because the Brownian motion $B(r^*)$ is Gaussian, the linear stochastic differential operator transforms a Gaussian process to another Gaussian process.

    In a similar way, we calculate the probability density function of fractional Brownian motion. Combining Equations \eqref{Schwarzschild-fBrownian} and \eqref{fFK}, we obtain
          \begin{equation*}
              \frac{\partial f(t,r^*)}{\partial r^*}=-\Big(1+\frac{2M}{M-r^*}\Big)\frac{\partial f(t,r^*)}{\partial t}+D H{r^*}^{2H-1}\frac{\partial^2 f(t,r^*)}{\partial t^2}.
          \end{equation*}

    The following proposition can be proved in the same way.
    \begin{prop}
    The probability density function of the stochastic radial motion driven by fractional Brownian motion of massless particles in Schwarzschild spacetime is
         	\begin{equation*}
         	    f(t,r^*) =\frac{1}{\sqrt{2\pi D({r^*}^{2H}-{r^*_0}^{2H})}}\exp \Big[-\frac{\big(t-\mu _{t(r^*)}\big)^2}
         {2\big(\sqrt{D({r^*}^{2H}-{r^*_0}^{2H})}\big)^2} \Big].    
         	\end{equation*}
    \end{prop}

    The probability $P$ of the stochastic process with respect to the probability density function $f$ is
	\begin{equation*}
	    P(x \leq X) = \int_{-\infty}^{x}f(x) \, dx.
	\end{equation*}

    Now we prove the first main theorem.
    
       \begin{thm}
        In Schwarzschild spacetime, the probability is zero for massless particles in the exterior region to reach the event horizon $r=2M$ in any finite coordinate time $T$ after adding Brownian motion or fractional Brownian motion. 
        
        Moreover, the probability is independent of the diffusion coefficient of Brownian motion $D$, Hurst parameter $H$ and the initial position $(t_0,r^*_0)$.
       \end{thm}
        \pf  
        \begin{equation*}
            \begin{aligned}
    P\{t<t',r^*|t_0,r^*_0\} & = \int_{0}^{t'}f(\tau,r^*|t_0,r^*_0) \, d\tau  \\
                       & = \frac{1}{\sqrt{2\pi}\sigma}\int_{0}^{t'} e^{ -\frac{(\tau-\mu)^2}{2\sigma ^2} }\, d\tau
        \end{aligned}
        \end{equation*}
    Without loss of generality, we set $t_0=0$, $r^*_0=0$,$r^*=M$. By the coordinate transform above, the black hole event horizon lies at $r^* =M$. For any finite time $T>0$,
\begin{equation*}
    \begin{aligned}
	P\{t<T,M|0,0\} & = \int_{0}^{T}f(\tau,M|0,0) \, d\tau  \\
	& = \frac{1}{\sqrt{2\pi}\sqrt{2DM}}\int_{0}^{T} e^{ -\frac{(\tau+M-2M\ln\frac {0}{C} )^2}{4DM}} \, d\tau \\
	& = 0.
\end{aligned}
\end{equation*}
The fractional Brownian motion case can be proved in the same way since its probability density function is also Gaussian and $\mu_{t(r^*)} \rightarrow -\infty$ as $r \rightarrow 2M$.
\qed

\mysection{Stochastic motion of massless particles in Reissner-Nordstr\"om spacetime}\ls

In this section, we calculate the probability for massless particles in the exterior region to reach the event horizon in finite coordinate time in Reissner-Nordstr\"om spacetime. We show that the probability is also zero. The method is almost the same as the Schwarzschild case, so we omit some details to avoid repetition.

We first consider the non-extreme Reissner-Nordstr\"om spacetime. The corresponding stochastic differential equation (SDE) for non-extreme Reissner-Nordstr\"om spacetime can be written as
\begin{equation*}
    \begin{aligned}
dt=-\Big(1+\frac{A}{r-r_+} + \frac{B}{r-r_-}\Big)dr+dB(r), \qquad  r_+ \leq r \leq 2M + \sqrt{M^2 -e^2},    
\end{aligned}
\end{equation*}
where $A,B,r_+,r_-$ are defined in Section 1. Take coordinate change
\begin{align*}
r^* = r_+ +M -r = 2M + \sqrt{M^2 -e^2} -r \in [0, M], 
\end{align*}
we get
\begin{equation}
    dt = \Big( 1 + \frac{A}{M - r^*} + \frac{B}{M + 2\sqrt{M^2 -e^2} -r^*} \Big)dr^* + dB(r^*).
    \label{nonextreme-B}
\end{equation}

\begin{prop}\label{p-nonextreme-Brownian} 
 	The probability density function of the stochastic radial motion driven by Brownian motion of massless particles in non-extreme $\text{Reissner-Nordstr\"om}$ spacetime is
 	\begin{equation*}
 	    f(t,r^*|t_0,r^*_0)  = \frac{1}{\sqrt{2\pi \sigma^2_{t(r^*)}}}
 		\exp\Big[-\frac{\big({t-\mu_{t(r^*)} }\big ) ^2}{2\sigma^2_{t(r^*)}}\Big],
 	\end{equation*}
 	where
    \begin{equation*}
 	\mu_{t(r^*)} = t_0- (r^*-r^*_0) + A \ln\frac{r^* - M}{r_0^* -M} + B\ln\frac{r^* -M -2\sqrt{M^2 -e^2}}{r_0^* -M -2\sqrt{M^2-e^2}}, 
    \end{equation*}
    \begin{equation*}
 	\sigma^2_{t(r^*)} =  2D (r^*-r^*_0).
    \end{equation*}
\end{prop}

\pf 
We give a sketch of proof and omit some repeated calculations.
Combining Eqs \eqref{nonextreme-B} and \eqref{FK}, the probability density function of stochastic radial motion in non-extreme Reissner-Nordstr\"om spacetime satisfies
\begin{equation}
    \frac{\partial f}{\partial r^*} = - \Big( 1 + \frac{A}{M-r^*} + \frac{B}{M + 2\sqrt{M^2 -e^2} - r^*}\Big)  \frac{\partial f}{\partial t} + D \frac{\partial^2 f}{\partial t^2}.
    \label{nonextreme-density}
\end{equation}
Consider the characteristic function
\begin{equation*}
    \phi(u, r^* \mid t_0, r_0) = \int_{-\infty}^{+\infty} e^{iut} f \, dt.
\end{equation*}
Taking Fourier transform on both sides of \eqref{nonextreme-density},
\begin{equation*}
    \frac{\partial \phi}{\partial r^*} = - \Big( 1 + \frac{A}{M-r^*} + \frac{B}{M + 2\sqrt{M^2 -e^2} - r^*}\Big) iu \phi - Du^2 \phi.
\end{equation*}
Solving the above equation with the initial condition $\phi = e^{iut_0}$,
\begin{equation*}
    \begin{aligned}
        \phi = \exp\Big\{ &\Big[t_0 - (r^* -r_0^*) + A\ln\frac{r^* -M}{r_0^* -M} + B \ln\frac{r^* -M -2\sqrt{M^2-e^2}}{r_0^* -M -2\sqrt{M^2 -e^2}} \Big]iu \\ 
        &- Du^2 (r^* -r_0^*) \Big\} \\
    \end{aligned}
\end{equation*}
Finally taking inverse transform, we get the probability density function as desired.
\qed

\begin{rmk}
    One can easily check that if $e = 0$, the probability density function of non-extreme Reissner-Nordstr\"om spacetime goes back to Schwarzschild case in proposition \ref{p-Schwarzshild-Brownian}.
\end{rmk}

Similarly, we calculate the probability density function of fractional Brownian motion in non-extreme $\text{Reissner-Nordstr\"om}$ spacetime. Comparing to Eq \eqref{fFK}, we get
\begin{equation}
    \frac{\partial f}{\partial r^*} = - \Big( 1 + \frac{A}{M-r^*} + \frac{B}{M + 2\sqrt{M^2 -e^2} - r^*}\Big)  \frac{\partial f}{\partial t} + DHr^{*2H-1} \frac{\partial^2 f}{\partial t^2}
    \label{nonextreme-fdensity}
\end{equation}

Thus we obtain
\begin{prop} \label{p-nonextreme-fB}
    The probability density function of the stochastic radial motion driven by fractional Brownian motion of massless particles in non-extreme $\text{Reissner-Nordstr\"om}$ spacetime is
         	\begin{equation*}
         	    f(t,r^*) =\frac{1}{\sqrt{2\pi D({r^*}^{2H}-{r^*_0}^{2H})}}\exp \Big[-\frac{\big(t-\mu _{t(r^*)}\big)^2}
         {2\big(\sqrt{D({r^*}^{2H}-{r^*_0}^{2H})}\big)^2} \Big],  
         	\end{equation*}
    where $\mu_{t(r^*)}$ same as proposition \ref{p-nonextreme-Brownian}.
\end{prop}

Next, we consider the extreme Reissner-Nordstr\"om spacetime. The corresponding stochastic differential equation (SDE) for extreme Reissner-Nordstr\"om spacetime can be written as
\begin{equation*}
    \begin{aligned}
dt=-\Big(1+\frac{2M}{r-M} + \frac{M^2}{(r-M)^2}\Big)dr+dB(r), \qquad M \leq r \leq 2M.     
\end{aligned}
\end{equation*}
Take coordinate change 
\begin{align*}
r^* = 2M - r \in [0, M], 
\end{align*}
we get
\begin{equation}
    dt = \Big( 1 + \frac{2M}{M - r^*} + \frac{M^2}{(M -r^*)^2} \Big)dr^* + dB(r^*).
    \label{extreme-B}
\end{equation}
Combining Eqs \eqref{extreme-B} and \eqref{FK}, the probability density function of the stochastic radial motion in extreme Reissner-Nordstr\"om spacetime satisfies
\begin{equation}
    \frac{\partial f}{\partial r^*} = - \Big( 1 + \frac{2M}{M-r^*} + \frac{M^2}{(M - r^*)^2}\Big)  \frac{\partial f}{\partial t} + D \frac{\partial^2 f}{\partial t^2}
    \label{extreme-density}
\end{equation}

\begin{prop} \label{p-extreme-B}
    The probability density function of the stochastic radial motion driven by Brownian motion of massless particles in extreme Reissner-Nordstr\"om spacetime is
    \begin{equation*}
 	    f(t,r^*|t_0,r^*_0)  = \frac{1}{\sqrt{2\pi \sigma^2_{t(r^*)}}}
 		\exp\Big[-\frac{\big({t-\mu_{t(r^*)} }\big ) ^2}{2\sigma^2_{t(r^*)}}\Big],
 	\end{equation*}
 	where
    \begin{equation*}
 	\begin{aligned}
 	    &\mu_{t(r^*)} = t_0- (r^* - r_0^*) + \Big( \frac{M^2}{r^* -M} - \frac{M^2}{r_0^*-M}\Big) + 2M\ln \frac{r^*-M}{r_0^*-M}, \\
        &\sigma^2_{t(r^*)} =  2D (r^*-r^*_0).
 	\end{aligned}
    \end{equation*}
\end{prop}

\begin{rmk}
    It is worth noting that if we pass the limit $M \rightarrow |e|$ or simply take $M = |e|$ in the non-extreme case (see proposition \ref{p-nonextreme-Brownian}), we \textbf{cannot} get the probability density function of extreme Reissner-Nordstr\"om spacetime above, even if we further assume that the coefficient identity $A+B=2M$ still holds.
    
    Mathematically speaking, the coefficients $A \rightarrow + \infty$ and $ B\rightarrow -\infty$ when $M \rightarrow |e|$, which makes the whole computation invalid. This reflects that the distinction between extreme and non-extreme black hole is preserved under Brownian perturbation. Similar phenomenon also occurs in the fractional Brownian motion case below.
\end{rmk}

    Following a similar derivation, we also obtain
\begin{prop} \label{p-extreme-fB}
    The probability density function of the stochastic radial motion driven by fractional Brownian motion of massless particles in extreme Reissner-Nordstr\"om spacetime is
         	\begin{equation*}
         	    f(t,r^*) =\frac{1}{\sqrt{2\pi D({r^*}^{2H}-{r^*_0}^{2H})}}\exp \Big[-\frac{\big(t-\mu _{t(r^*)}\big)^2}
         {2\big(\sqrt{D({r^*}^{2H}-{r^*_0}^{2H})}\big)^2} \Big],  
         	\end{equation*}
    where $\mu_{t(r^*)}$ same as proposition \ref{p-extreme-B}.
\end{prop}

Based on propositions \ref{p-nonextreme-Brownian},\ \ref{p-nonextreme-fB},\ \ref{p-extreme-B}, and \ref{p-extreme-fB}, we obtain the second main theorem.
\begin{thm}
    In Reissner-Nordstr\"om spacetime, the probability is zero for massless particles in the exterior region to reach the event horizon ($r_+ = M + \sqrt{M^2-e^2}$ for the non-extreme case, $r=M$ for the extreme case) in any finite coordinate time $T$ after adding Brownian motion or fractional Brownian motion.
    
    Moreover, the probability is independent of the diffusion coefficient of Brownian motion $D$, Hurst parameter $H$ and the initial position $(t_0,r^*_0)$.
\end{thm}

\pf
We utilize that the probability density function is Gaussian and that $\mu_{t(r^*)} \rightarrow -\infty$ as $r$ approximates the event horizon. Then by a direct integration of the probability density functions listed above, we obtain that the probability is zero.
\qed

\mysection{Conclusion}\ls

In Schwarzschild and Reissner-Nordstr\"om spacetimes, classically, massless particles in the exterior region take infinite coordinate time to reach the event horizons. By adding noise kernels as the quantum fluctuations of gravity and replacing the semiclassical Einstein equations by the Einstein-Langevin equations, we show that for any given finite time, the probability for massless particles reaching the event horizons is still zero. For Reissner-Nordstr\"om spacetime, we further discover that the distinction between extreme and non-extreme black hole is also preserved. These stochastic perturbations preserve certain properties of classical black holes.

\bigskip

{\footnotesize {\it Acknowledgement. The first four sections of this paper are partially based on the second author's master's thesis at Guangxi University. The first author was supported by Beijing Natural Science Foundation (Grant No. QY24355). The third author was supported by the special foundation for Guangxi Ba Gui Scholars. The authors gratefully acknowledge the hospitality of Guangxi Center for Mathematical Research, Guangxi University, where the idea for this paper originated.}

\end{document}